\documentclass[11pt]{article}

\date{}

\usepackage[utf8]{inputenc} 
\usepackage[T1]{fontenc}    
\usepackage[colorlinks=true,linkcolor=blue,allcolors=blue]{hyperref}       
\usepackage{url}            
\usepackage{booktabs}       
\usepackage{amsfonts}       
\usepackage{nicefrac}       
\usepackage{microtype,nicefrac}      
\usepackage{xspace}
\usepackage{natbib}

\usepackage{comment}
\usepackage{amsmath}
\usepackage{graphicx}
\usepackage{rotating}

\usepackage{amsmath}
\usepackage{amssymb}
\usepackage{autobreak}
\usepackage{mathtools}
\usepackage[dvipsnames]{xcolor}
\usepackage{bbm}
\usepackage[ruled,vlined, linesnumbered]{algorithm2e}
\usepackage{algorithmic}
\usepackage[parfill]{parskip}

\usepackage[toc]{appendix}
\usepackage{minitoc}
\usepackage{caption}
\usepackage{subcaption}

\makeatletter

\DeclareMathOperator*{\argmax}{arg\,max}
\DeclareMathOperator*{\argmin}{arg\,min}

\newcommand{\N}{\mathcal N}

\title{Neural mixture model association of seismic phases}

\author{%
  Zachary E. Ross\\ 
  Seismological Laboratory\\
  California Institute of Technology\\
  \texttt{zross@caltech.edu} \\
  \and
  Weiqiang Zhu\\ 
  Seismological Laboratory\\
  California Institute of Technology\\
  \texttt{wqzhu@caltech.edu} \\
  \and
  Kamyar Azizzadenesheli\\ 
  Nvidia Corp.\\
  \texttt{kamyara@nvidia.com}
}

\begin{document}
\doparttoc 
\faketableofcontents 

\maketitle

\begin{abstract}
Seismic phase association is the task of grouping phase arrival picks across a seismic network into subsets with common origins. Building on recent successes in this area with machine learning tools, we introduce a neural mixture model association algorithm (Neuma), which incorporates physics-informed neural networks and mixture models to address this challenging problem. Our formulation assumes explicitly that a dataset contains real phase picks from earthquakes and noise picks resulting from phase picking mistakes and fake picks. The problem statement is then to assign each observation to either an earthquake or noise. We iteratively update a set of hypocenters and magnitudes while determining the optimal class assignment for each pick. We show that by using a physics-informed Eikonal solver as the forward model, we can impose stringent quality control on surviving picks while maintaining high recall. We evaluate the performance of Neuma against several baseline algorithms on a series of challenging synthetic datasets and the 2019 Ridgecrest, California sequence. Neuma outperforms the baselines in precision and recall for each of the synthetic datasets. Furthermore, it detects an additional 3285 more earthquakes than the best baseline on the Ridgecrest dataset (13.5\%), while substantially improving the hypocenters.
\end{abstract}
\textit{Keywords:
}

\section{Introduction}
Seismicity catalogs are the foundation for a wide array of seismological analyses. They are generally produced in automated fashion with detection and location algorithms, which include phase picking algorithms and array processing methods. The phase picking approach is the one adopted most commonly by real-time seismic networks, and consists of two main steps. First, the raw continuous waveform data is processed by phase detection/picking algorithms, one station at a time, to identify candidate earthquake signals and measure their onset times. Then, a second algorithm is applied to consider combinations of these detections across the network of sensors and decide whether any disjoint subsets could originate from coherent sources. This step is referred to as seismic phase association and is responsible for determining which of the candidate phase picks should be used to locate the event and determine its magnitude. For large magnitude events with high signal-to-noise ratios, phase association is fairly straightforward, as the number of phase picks across the network is generally large. Detecting small earthquakes, particularly during aftershock sequences and swarms, is quite challenging however due to relatively fewer stations having picks available, the presence of lots of noise picks, and overprinting of seismic wave moveouts due to nearly concurrent events \citep[e.g.][]{ross_phaselink:_2019}.

For many years, phase association algorithms were largely based around grid search algorithms, in which phase arrival times were back-projected to see if any subset focused at a coherent origin time \citep{le_bras_global_1994,draelos_new_2015,patton_hydranational_2016,yeck_glass3_2019}. These methods are strictly based on wave propagation physics. More recently, there have been several phase association algorithms that incorporate machine learning or more modern elements from applied mathematics, \citep{reynen_supervised_2017,mcbrearty_earthquake_2019,ross_phaselink:_2019,zhang_rapid_2019,woollam_hex_2020,zhu_earthquake_2022}. These methods all incorporate aspects of the wave physics (either explicitly or implicitly) in solving the association problem, but are wide-ranging in the types of algorithms employed to do so. Of these, the Gaussian Mixture Model Association algorithm \citep[GaMMA;][]{zhu_earthquake_2022} is particularly appealing because it is able to not only use travel time information, but also amplitude information, to better constrain the results. GaMMA formulates the association problem as one of unsupervised (probabilistic) clustering, where the centroids of the clusters are hypocenter-magnitude pairs and the error in the picks and forward model is captured by a Gaussian distribution; the problem statement then is to optimize the hypocenters while simultaneously grouping together phase picks that match them.

GaMMA was shown to perform especially well on the very active 2019 Ridgecrest, California earthquake sequence \citep{zhu_earthquake_2022}. However, the method has a number of shortcomings that hinder its potential. One clear limitation is that the technique uses a forward model for the travel times that is based on a homogeneous velocity structure, resulting in travel times that can be substantially inaccurate. \citet{zhu_earthquake_2022} overcome this in part by using a very large (2.0 s) uncertainty on the travel times, but this comes with the cost of potentially including noise picks too. Another significant limitation of GaMMA is that it assumes that all picks were produced by earthquakes when fitting the model, even if in reality a pick was a false detection. As we have strong \textit{a priori} knowledge that a considerable fraction of the input picks will be noise picks output from the phase pickers \citep[e.g][]{ross_generalized_2018}, it would be ideal to include this knowledge explicitly into a model.

In this study, we introduce a Neural mixture model association algorithm, Neuma. Our contributions are three-fold: (i) We modify the Gaussian mixture model association algorithm to use a neural network Eikonal solver, eliminating the need for a homogeneous velocity model and leading to considerably improved phase associations, more detections, and more precise hypocenters. (ii) We introduce a new probability formulation that includes latent noise variables in addition to real picks, allowing for explicit noise labels to be assigned. (iii) Finally, we introduce a separate warmup scheme for initializing the hypocenters that helps to achieve better global convergence. We demonstrate the performance of our method on several synthetic datasets and the 2019 Ridgecrest, California sequence and compare the results to several established baselines. For the Ridgecrest dataset, we detect an additional 3283 earthquakes over the best baseline, an improvement of 13.4\%, while reducing the number of associated picks by 2 per event on average. In addition, we reducethe bias in hypocentral depth by nearly 5~km per event on average.
 
\section{Method}\label{Sec:SV}
\subsection{Preliminaries}

Seismic phase association is the problem of organizing a set of tentative phase detections from different sensors into subsets that have common hypocenters. Let $y=\{(t_i, a_i)\}_{i=1}^N$ denote a set of observed phase arrival picks and corresponding peak amplitudes. In most real-world cases, some of these picks will correspond to earthquakes, whereas the remainder will be noise that needs to be filtered out. The problem statement is therefore to assign each $y_i$ to one of $K+1$ classes, where $K$ is the number of earthquakes and the $(K+1)$th class represents noise. An earthquake is defined by a hypocenter $x_i \in \mathbb{R}^4$ (i.e. location and origin time) and magnitude $M_i \in \mathbb{R}$, and the observations $y$ occur over some time interval $\Delta T$. Furthermore, let $M$ denote the set of all magnitudes, i.e., $M:=\{M_i\}_i^N$ and $x$ denote the set of all hypocenters, i.e., $x:=\{x_i\}_i^{N}$.

If the probability of an observation belonging to class $k$ is $\phi_k$, and $\phi$ denotes the set of all event probabilities, i.e., $\phi:=\{\phi_k\}_k^{K+1}$, then the joint likelihood of the set of observations can be written as,
\begin{equation}
    p(t, a |\phi,x,M) =  \prod_i^N p(t_i, a_i |\phi,x,M).
\end{equation}
For a mixture model with $K+1$ classes, this (log)-likelihood can be written as,
\begin{equation}\label{eq:lik}
    \log p(t,a|\phi,x,M) = \sum_{i=1}^N \log \left[\sum_{k=1}^K \phi_k p(t_i, a_i | x_k, M_k) + \phi_{K+1}p^G(t_i, a_i)\right].
\end{equation}
where $p^G(t_i, a_i)=p_T^G(t_i)p_A^G(a_i)$ is a joint density for noise picks. 

\begin{figure}[htb]
    \centering
    \includegraphics[width=0.6\textwidth]{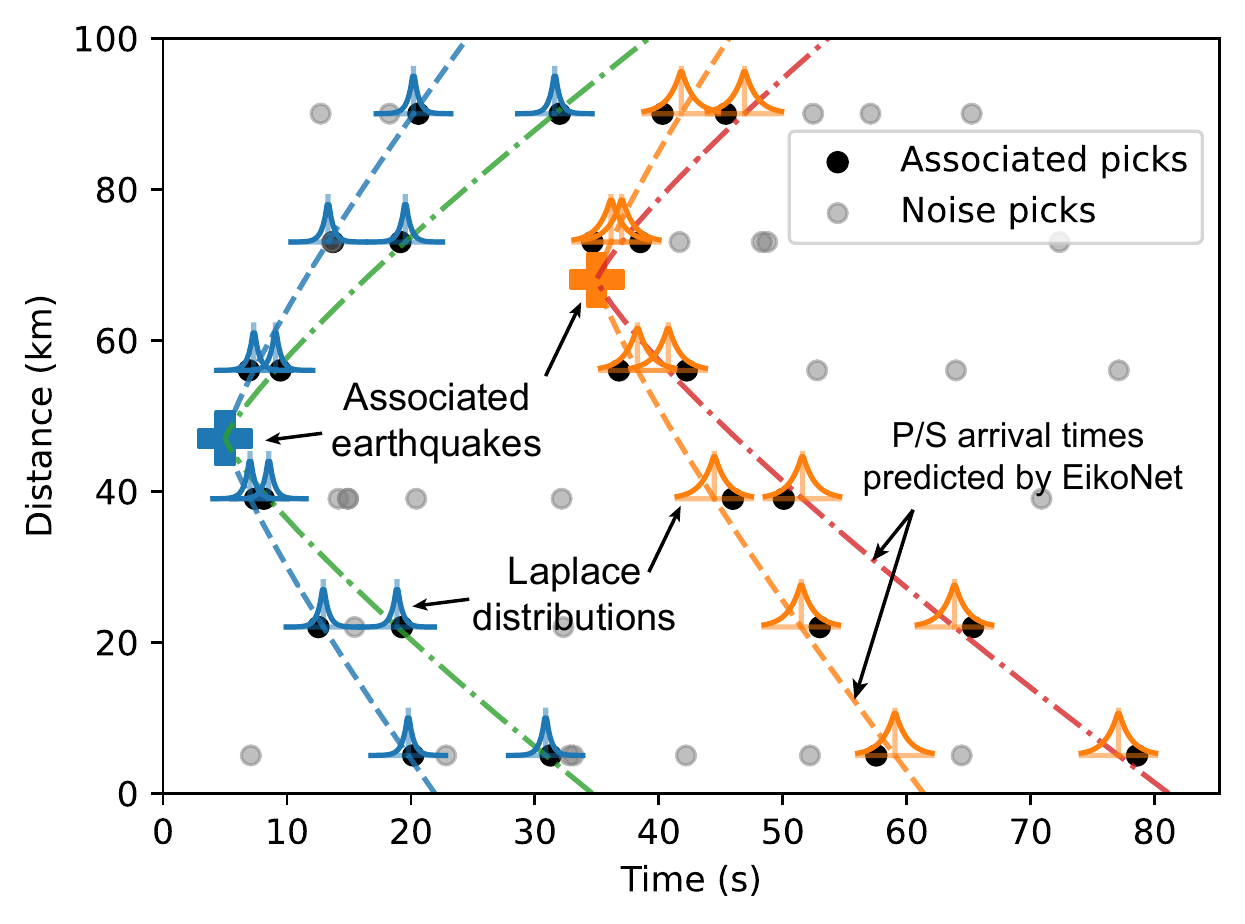}
    \caption{A cartoon depicting neural mixture model association (Neuma). A set of initially disordered phase picks and amplitudes are iteratively grouped into disjoint subsets that have common hypocenters and magnitudes.}
    \label{fig:cartoon}
\end{figure}

Neuma is schematically illustrated in Figure \ref{fig:cartoon}. We are given an initially unclustered set of phase picks and amplitudes. We are tasked with identifying which of the picks should be labeled as noise, and which of the picks should be assigned to disjoint subsets that correspond to individual earthquakes. The subset of picks associated to each earthquake is described by a forward model and a probability model for the observation uncertainty (pick and amplitude error). 

\paragraph{Expectation-Maximization.} Optimizing the likelihood in eq. \ref{eq:lik} is complicated by the summation inside the logarithm, which results from the class assignments for each observation being unknown. One of the most common approaches to maximizing likelihoods with latent variables is expectation-maximization \citep[E-M;][]{dempster1977maximum}, in which the class assignments are first computed under the fixed probability model (the "E" step) and then the cluster parameters are optimized using these newly determined class assignments (the "M" step). By iterating between the E-step and M-step, the E-M algorithm has provable non-decreasing convergence, making it appealing for solving likelihood problems with latent variables.

\paragraph{E-step.}
Let $\gamma_{ik}$ denote the probability of the $i$th phase pick belonging to class $k$. Given a set of hypocenters and magnitudes, we can compute $\gamma$ to update $\phi$ as follows,

\begin{equation}
    \gamma_{i,k} = \frac{\phi_k p_T^E(t_i | x_k, \theta^T) p_A^E(a_i | x_k, M_k,  \theta^A)}{
    {\sum_k^K\phi_k p_T^E(t_i | x_k, \theta^T) p_A^E(a_i | x_k, M_k,  \theta^A)+\phi_{K+1} p_T^G(t^i)p_A^G(a_i)}}
\end{equation}
for all $k\in[K]$, and 
\begin{equation}
    \gamma_{i,K+1} = \frac{\phi_{K+1} p_T^G(t^i)p_A^G(a_i)}{
    \sum_k^K\phi_k p_T^E(t_i | x_k, \theta^T) p_A^E(a_i | x_k, M_k,  \theta^A)+\phi_{K+1} p_T^G(t^i)p_A^G(a_i)}
\end{equation}



Therefore, $\forall k\in[K+1]$ we have, 

\begin{equation*}
    \phi_k=\frac{\sum_{i=1}^N\gamma_{i,k}}{N}.
\end{equation*}

\paragraph{M-step.}

The maximization step of the E-M algorithm uses the newly determined class assignments from the E-step to compute the expectation of the likelihood in eq. \ref{eq:lik} and then maximize it. Mathematically this corresponds to,

\begin{equation*}
    x,M=\argmax_{x,M}\log p(y|\phi,x,M).
\end{equation*}

The $\gamma_{ik}$ from the E-step are thus used to update $x$ and $M$,

\begin{equation}
x_k = \argmin_{x_k} \sum_{i=1}^N \gamma_{ik} | t_i - t(x_i, x_k)|
\end{equation}

\begin{equation}
M_k = \frac{\sum_{i=1}^N \gamma_{ik} M(x_k, x_i, a_i)}{\sum_{i=1}^N\gamma_{ik}}
\end{equation}

where $M(x_k, x_i, a_i)$ is an empirical magnitude definition e.g. local magnitude. Here, we use the same equation as \citet{zhu_earthquake_2022}, taken from \citet{picozzi_rapid_2018},

\begin{equation*}
    \log_{10} PGV = 1.08 + 0.93(M-3.5) - 1.68\log_{10} R
\end{equation*}
where $PGV$ is peak ground velocity in units of cm/s, $M$ is magnitude, and $R$ is hypocentral distance in km.

\subsection{Implementation}

\paragraph{Choice of density functions.}

The densities $p_T^E$, $p_A^E$, $p_T^G$, and $p_A^G$ define the arrival time and amplitude likelihoods for the earthquake and noise classes. The earthquake densities were taken to be Gaussian by \citet{zhu_earthquake_2022}, with
\begin{equation*}
    p_T^E(t_i | x_k, \theta^T)=\N(t(x_k, x_i), \Lambda_k^T)
\end{equation*}
\begin{equation*}
    p_T^A(a_i | x_k, M_k, \theta^A)=\N(a(x_k, x_i, M_k), \Lambda_k^A),
\end{equation*}
where $t(x_k, x_i)$ and $a(x_k, x_i, M_k)$ are forward models for arrival times and amplitudes for a source with hypocenter $x_k$ at receiver $x_i$ with magnitude $M_k$. For this study, we instead use Laplace distributions, 
\begin{equation*}
p_T^E(t_i | x_k, \theta^T) = \text{Laplace}(\mu=t(x_k, x_i), b^T)
\end{equation*}
\begin{equation*}
p_T^A(a_i | x_k, M_k, \theta^A) = \text{Laplace}(\mu=a(x_k, x_i, M_k), b^A).
\end{equation*}

The scale parameters $b^T \in \mathbb{R}$ and $b^A \in \mathbb{R}$ in these distributions are taken to represent the combined uncertain from our measurement error and forward model. The use of a Laplace distribution over a Gaussian distribution is motivated by the observations that arrival time residuals made with deep learning algorithms are closer to a Laplace distribution than a Normal distribution \citep[e.g.][]{ross_p_2018,zhu_phasenet:_2019}. Additionally, the $\ell^1$ norm in the Laplace distribution is more robust against outliers, which are common in phase picking datasets.

\begin{figure}[htb]
    \centering
    \includegraphics[width=0.5\textwidth]{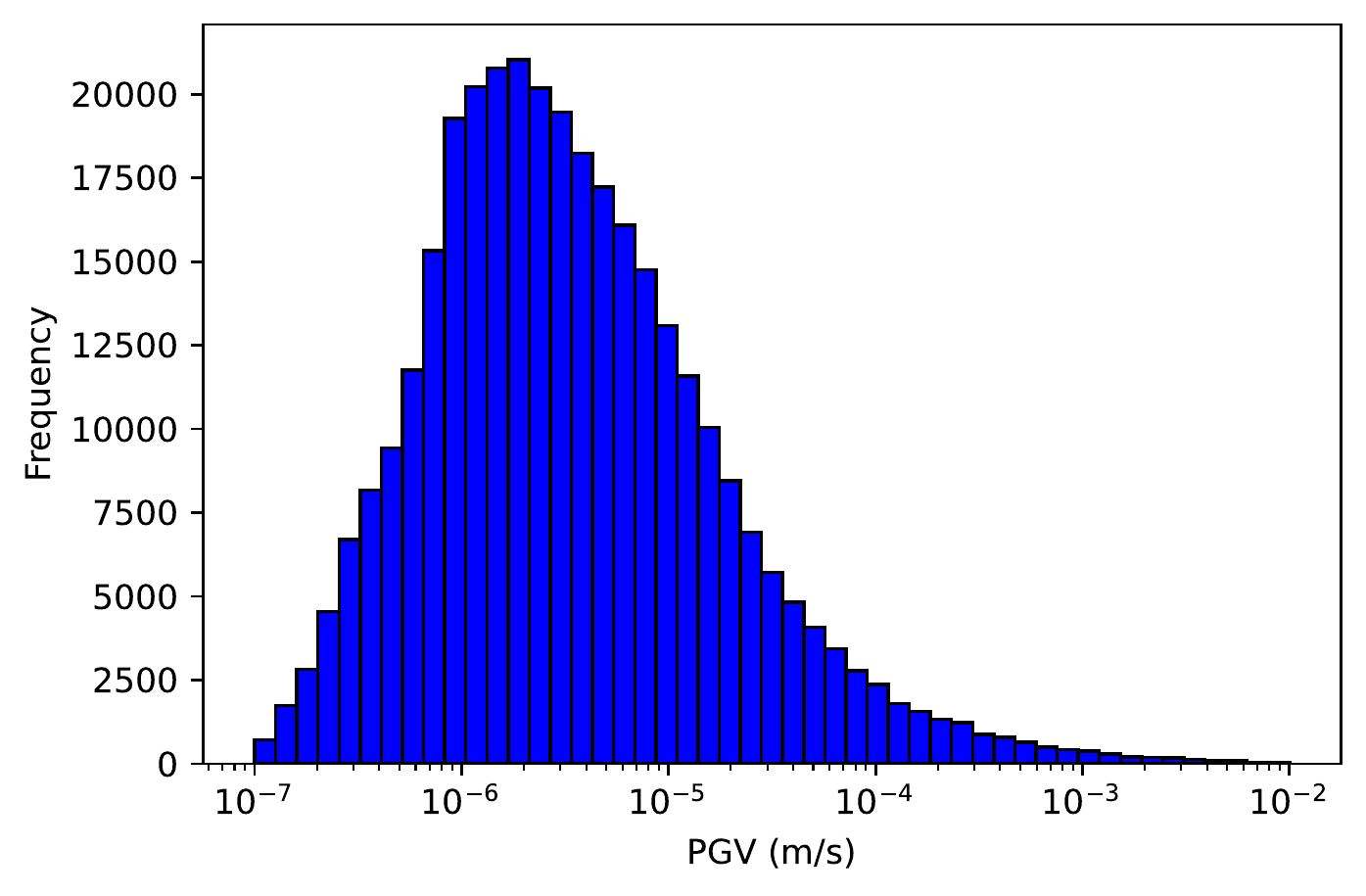}
    \caption{noise pick amplitude distribution.}
    \label{fig:garbageampdist}
\end{figure}

We also need the densities $p_T^G$, and $p_A^G$ for the noise. For the arrival time noise distribution, we assume that $p_T^G(t)$ is a uniform distribution on $\Delta t$. The noise amplitude distribution, $p_A^G$, does not have any strong theoretical motivation, particularly at the high frequencies of interest in this work. As a result we model this distribution empirically using the distribution of amplitudes for picks rejected as earthquakes by GaMMA in the Ridgecrest dataset produced in \citet{zhu_earthquake_2022}. This distribution is shown in Fig. \ref{fig:garbageampdist} and is approximately log-normal, with best-fitting parameters $\mu=-5.46$ and $\sigma=0.72$ in $\log_{10} m/s$ units. We use these parameters throughout this study. While there is the possibility that some of these amplitudes actually correspond to earthquakes that were missed by the associator, the majority are likely to be noise. 

\paragraph{Forward Models.}
The forward model used to compute $\mu_k$ for Neuma is an EikoNet \citep{smith_eikonet_2020}, a physics-informed deep neural network trained to solve the Eikonal equation for ray tracing. We use a simplified network architecture to maximize computational efficiency, with 5 dense layers of 128 neurons followed by exponential linear unit activations. We perform the M-step update of the $K$ hypocenters using the Adam optimizer \citep{kingma_adam:_2014} in a non-stochastic (all data points used) fashion.

\paragraph{Warmup Iterations.}
Our implementation of the E-M equations includes a few changes from that described by \citet{zhu_earthquake_2022}. We found that if the uncertainties on the likelihood functions, $b^T$ and $b^A$, were set from iteration 1 onward to values representative of the true velocity model and picking errors, that the E-M would have poor convergence and assign most picks automatically to the noise class. The reason for this is because if the initial hypocenters assigned to the $K$ clusters are not sufficiently close to the true position, that the E-step assignments for $\gamma$ would always start out being effectively zero for all picks, which is lower than the probability of the noise distribution. To circumvent this problem, we designed an implementation that instead has a warmup period in which the noise variables are turned off (i.e. $\gamma_{i,K+1}=0$) to allow for the $K$ hypocenters to first be optimized. During this period, we also decay the numerical values of $\Lambda_k$ for both the travel times and the amplitudes. At iteration 1, $\Lambda_k$ is set to 5 times the actual (assumed) uncertainty of the travel times and amplitudes, and then is decayed linearly to the actual uncertainty over 10 warmup iterations. The initial value of the uncertainty was chosen as a compromise being much larger than the true (final) uncertainty, but not so large that all clusters have the same likelihood of pick assignment. After the warmup period is over, $\gamma_{i,K+1}$ is computed at each iteration rather than being forced to zero, and $\Lambda_k$ is set to the constant values taken at the final iteration of the warmup. We found that the E-M converged typically within about 5 iterations after the warmup period was finished.

\paragraph{Selection of $K$.}
\citet{zhu_earthquake_2022} showed that the parameter $K$ has a major influence on the performance of the method, particularly in terms of computational efficiency. This is because the computational complexity scales non-linearly with $K$. If too low, however, then some earthquakes may be missed entirely. The computational performance also scales with the number of observations. These issues generally require that a dataset be broken up into smaller chunks, which has the additional advantage of being fully parallelizable. \citet{zhu_earthquake_2022} defined the oversample factor to help select $K$, in which $K$ is set to some integer multiple of the ratio between the total number of picks in an observation window to the total number of stations available. It was shown that choosing a value of $K$ that is larger than the number of earthquakes usually results in the presence of some clusters with few or no picks associated, making the main tradeoff in $K$ being poor computational performance if too large, and having false negative detections if too small. In this regard, Neuma is essentially unchanged from GaMMA and we use in this paper an oversample factor of 4.0.

\paragraph{Breaking up a set of picks into subsets.}
Limiting the value of $K$ to something computationally manageable also requires breaking up a larger dataset into smaller chunks. \citet{zhu_earthquake_2022} achieved this using the density-based clustering algorithm DBSCAN \citep{ester_density-based_1996} to cluster the picks in time. This requires some hyperparameter tuning to achieve clusters that are long enough in time that they capture entire earthquakes, but not so long that the clusters run on for more than a few minutes. We also use DBSCAN for this purpose and set the parameter $\epsilon=5.0$~s with a minimum of 1 pick to merge clusters. We discard clusters with fewer picks than the minimum number required for detection (details to come in the next section). These values will however vary strongly based on the dataset attributes including the number of stations and their geometry, the number of false picks, and so forth. Thus for any new dataset these hyperparameters will require some tuning.

\section{Experiments}\label{Sec:Exp}

\subsection{Baselines}
We use several different baselines in this section to evaluate the performance of Neuma. In addition to Neuma, we test three other association algorithms. PhaseLink \citep{ross_phaselink:_2019} is an association algorithm that uses deep recurrent neural networks trained in supervised fashion and incorporates an unsupervised cluster analysis to group picks into events. GaMMA \citep{zhu_earthquake_2022} is an approach based on Gaussian mixture models in which the centroids of the clusters are earthquakes. For the real dataset, we use a catalog produced by the SCSN that was manually reviewed \citep{scedc_data_2013}, which was produced with the Binder phase associator as part of the Earthworm package.

\subsection{Synthetic datasets}
\begin{figure}[htb]
    \centering
    \includegraphics[width=1.0\textwidth]{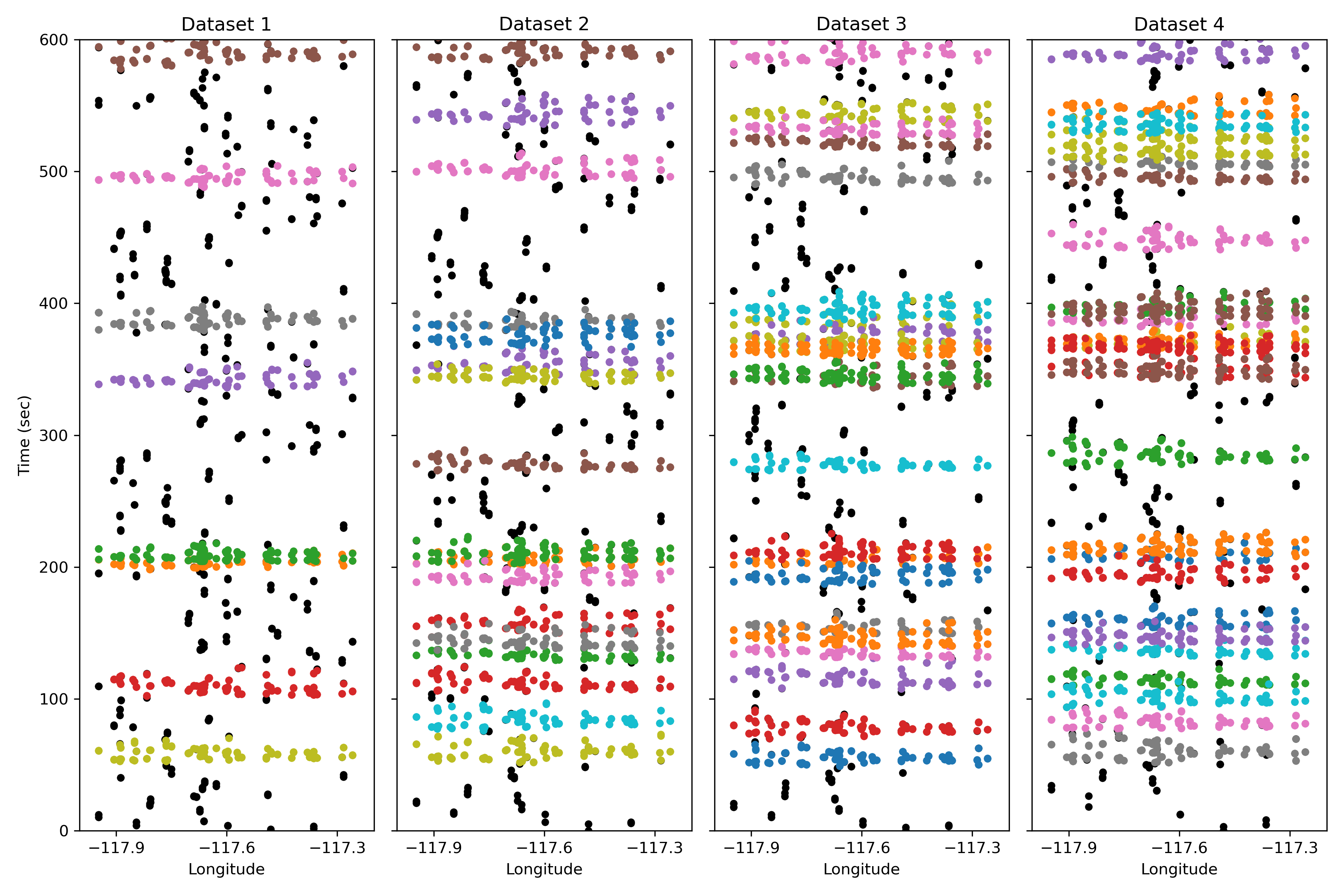}
    \caption{Samples of the synthetic datasets. Only the first 10 minutes of each dataset is shown. Colored points indicate real phase picks and are colored according to the associated event. Black points indicate noise picks.}
    \label{fig:syndata}
\end{figure}

For the first experiment, we generate a suite of four synthetic datasets of earthquakes (picks) that consist of both "true" earthquakes and garbage picks. The advantage of these datasets is that we have exact ground truth for every pick. We construct the datasets such that each one is increasingly more challenging than the last. We specify for each dataset a fixed number of earthquakes that are given random origin times and random uniform hypocenters. We then generate synthetic arrival times and amplitudes for the Southern California Seismic Network stations located within the study area \citep{scedc_data_2013}. We design each dataset such that it spans exactly 24 hours (Fig. \ref{fig:syndata}). The synthetic amplitudes are drawn from the empirical amplitude distribution defined in the previous section. The easiest dataset (D1) contains 1080 earthquakes in it, making the average time between earthquakes 80 seconds. The remaining datasets, D2, D3, and D4, have an average interevent time of 60, 40, and 20 seconds, respectively (Fig. \ref{fig:syndata}). We add 0.2 sec Gaussian noise to each arrival and 1.0 log10 unit of noise to the amplitudes; we set $b^A$ and $b^T$ to these values for the experiment. In addition, we create 57600 garbage picks with uniform times, randomly chosen stations and phases. Magnitudes for each event are set to a fixed value of 3.0 for these tests for the sake of simplicity.
To evaluate the performance of Neuma on these ground truth datasets, we use two metrics as introduced by \citet{ross_phaselink:_2019}. Since earthquake detections are really defined by a collection of phase picks, we use analogs of precision and recall that are more appropriate for collections of sets. To determine whether the $k$th cluster of picks, $A_k$, corresponds to a successful event detection, we define the precision between it and the best matching cluster of picks in the ground truth, $B_i$,
\begin{equation*}
P_k = \max_i\, N(A_k \cap B_i),\quad \forall i\in\{1,..,C\}\,,
\end{equation*}
where, $C$ is the total number of events (clusters) in the ground truth and $N(\cdot)$ denotes the cardinality of the set. We also define the recall as,
\begin{equation*}
R_i = \max_k\, N(A_k \cap B_i),\quad \forall k\in\{1,..,D\}\,,
\end{equation*}
where $D$ is the number of detected events (clusters). From these definitions, we can define the set-averaged precision, $P$, and recall, $R$, as,
\begin{equation*}
    P = \frac{\sum_k^D P_k}{\sum_k^D N(A_k)}
\end{equation*}
\begin{equation*}
    R = \frac{\sum_i^C R_i}{\sum_i^C N(B_i)}\,.
\end{equation*}
As with the classical definitions of precision and recall, $P$ describes how likely a single phase association is to be correct, whereas $R$ describes how likely each phase association in the ground truth is correctly recovered. These metrics naturally account for the potential for erroneously split and merged detections as both of these scenarios will impact the scores.

The precision and recall scores for each dataset are shown in Tables \ref{tab:prec}-\ref{tab:rec}. We can see that Neuma outperforms both baselines for all datasets in both precision and recall. Even for the hardest dataset, D4, Neuma achieves $P=0.952$, which is substantially better than the $P=0.901$ and $P=0.928$ for GaMMA and PhaseLink, respectively. The recall differences are even more stark, as Neuma achieves $R=0.947$ on D4, whereas GaMMA gets $R=0.743$ and PhaseLink only obtains $R=0.228$.

\begin{table}[]
\centering
\caption{Precision scores for the synthetic datasets}
\label{tab:prec}
\begin{tabular}{@{}lllll@{}}
\toprule
Algorithm  & D1    & D2    & D3    & D4 \\ \midrule
PhaseLink  & 0.976 & 0.961 & 0.942 & 0.928  \\
GaMMA      & 0.965 & 0.943 & 0.917 & 0.901  \\
Neuma & \textbf{0.979} & \textbf{0.975} & \textbf{0.965} & \textbf{0.952}  \\ \bottomrule
\end{tabular}
\end{table}


\begin{table}[]
\centering
\caption{Recall scores for the synthetic datasets}
\label{tab:rec}
\begin{tabular}{@{}lllll@{}}
\toprule
Algorithm  & D1    & D2    & D3    & D4 \\ \midrule
PhaseLink  & 0.664     & 0.530 & 0.427 & 0.228  \\
GaMMA      & 0.887 & 0.821 & 0.770 & 0.743  \\
Neuma & \textbf{0.989} & \textbf{0.977} & \textbf{0.955} & \textbf{0.947}  \\ \bottomrule
\end{tabular}
\end{table}

\subsection{Ridgecrest Dataset}

To test Neuma on an established real dataset, we apply it to the 2019 Ridgecrest earthquake sequence \citep{ross_hierarchical_2019} for the time period 2019-07-04 to 2019-07-09. This dataset was previously used as a benchmark by \citet{zhu_earthquake_2022} in tuning GaMMA, so it provides a clear comparison with a state of the art automated algorithm. This dataset has a manually reviewed seismicity catalog produced by the Southern California Seismic Network (SCSN), including phases \citep{scedc_data_2013}. Since this dataset does not have ground truth available, we use a variety of measures to evaluate the overall quality of the results, including magnitude distributions, temporal analysis, phase association counts, and location quality. Since PhaseLink does not compute magnitudes, we do not use it as a baseline for this dataset, and instead rely on the SCSN catalog which does have them. To facilitate analysis strictly at the association level, rather than the phase detection level, we use the same input set of phase detections between GaMMA and Neuma that were produced by \citet{zhu_earthquake_2022}; these were made by the PhaseNet algorithm \citep{zhu_phasenet:_2019}. It should be noted however that the SCSN results use moving average phase detectors.

\begin{figure}[htb]
    \centering
    \includegraphics[width=0.5\textwidth]{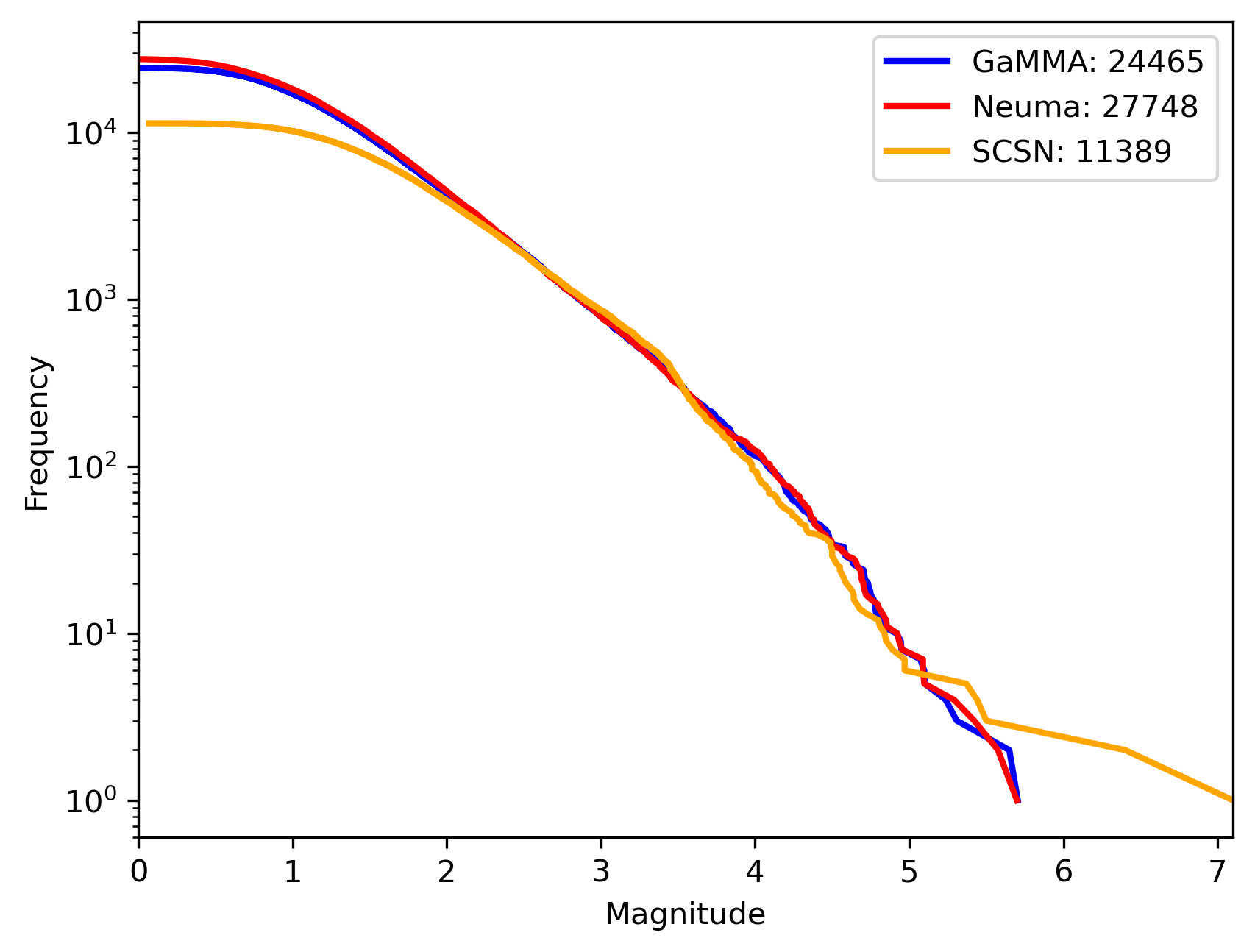}
    \caption{Ridgecrest frequency-magnitude statistics for the three algorithms.}
    \label{fig:rcfreqmag}
\end{figure}

\begin{figure}[ht]
    \begin{subfigure}[t]{0.5\columnwidth}
    \includegraphics[width=1.\textwidth]{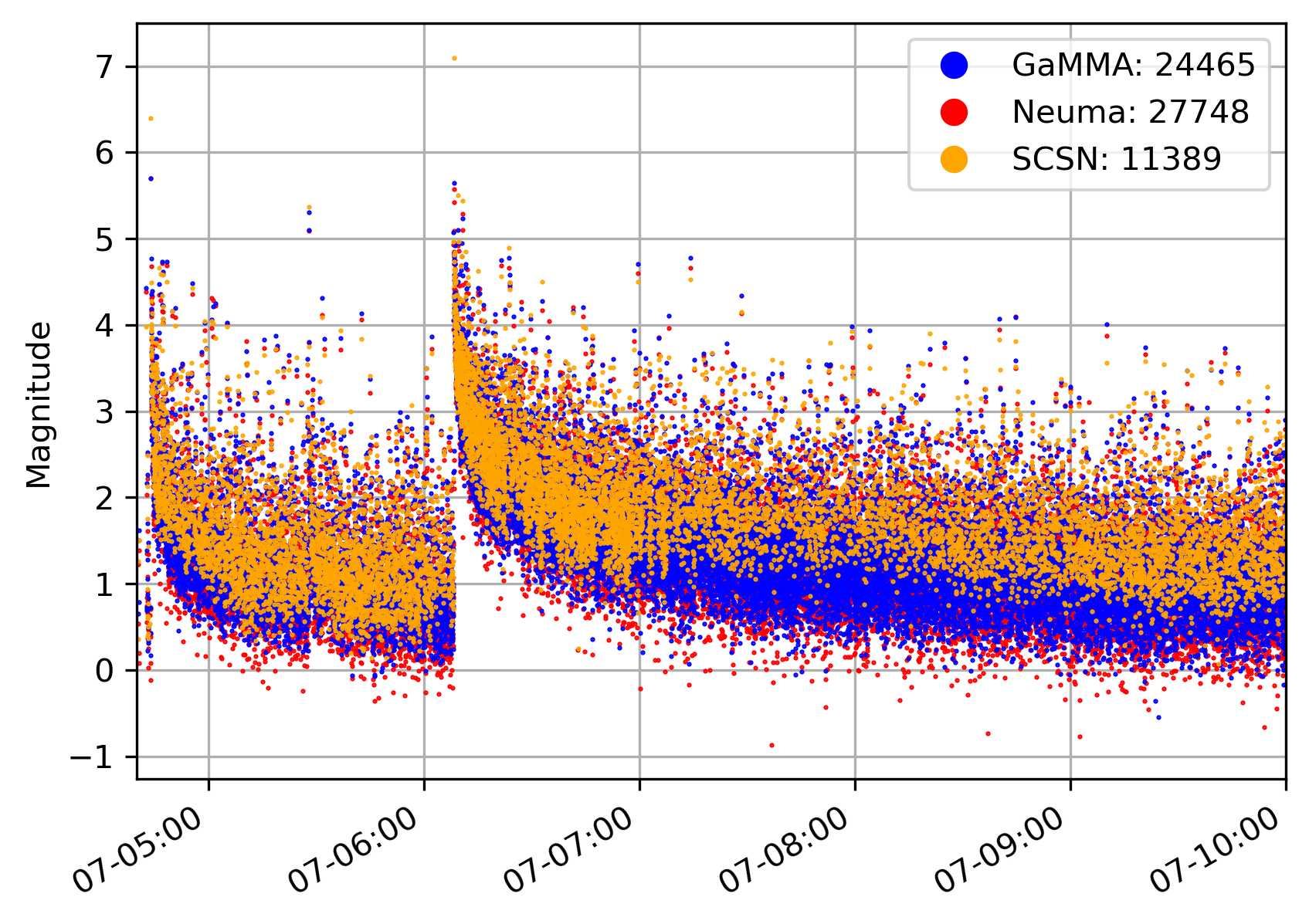}
    \label{fig:magtime}
    \end{subfigure}
    \begin{subfigure}[t]{0.5\columnwidth}
    \includegraphics[width=1.\textwidth]{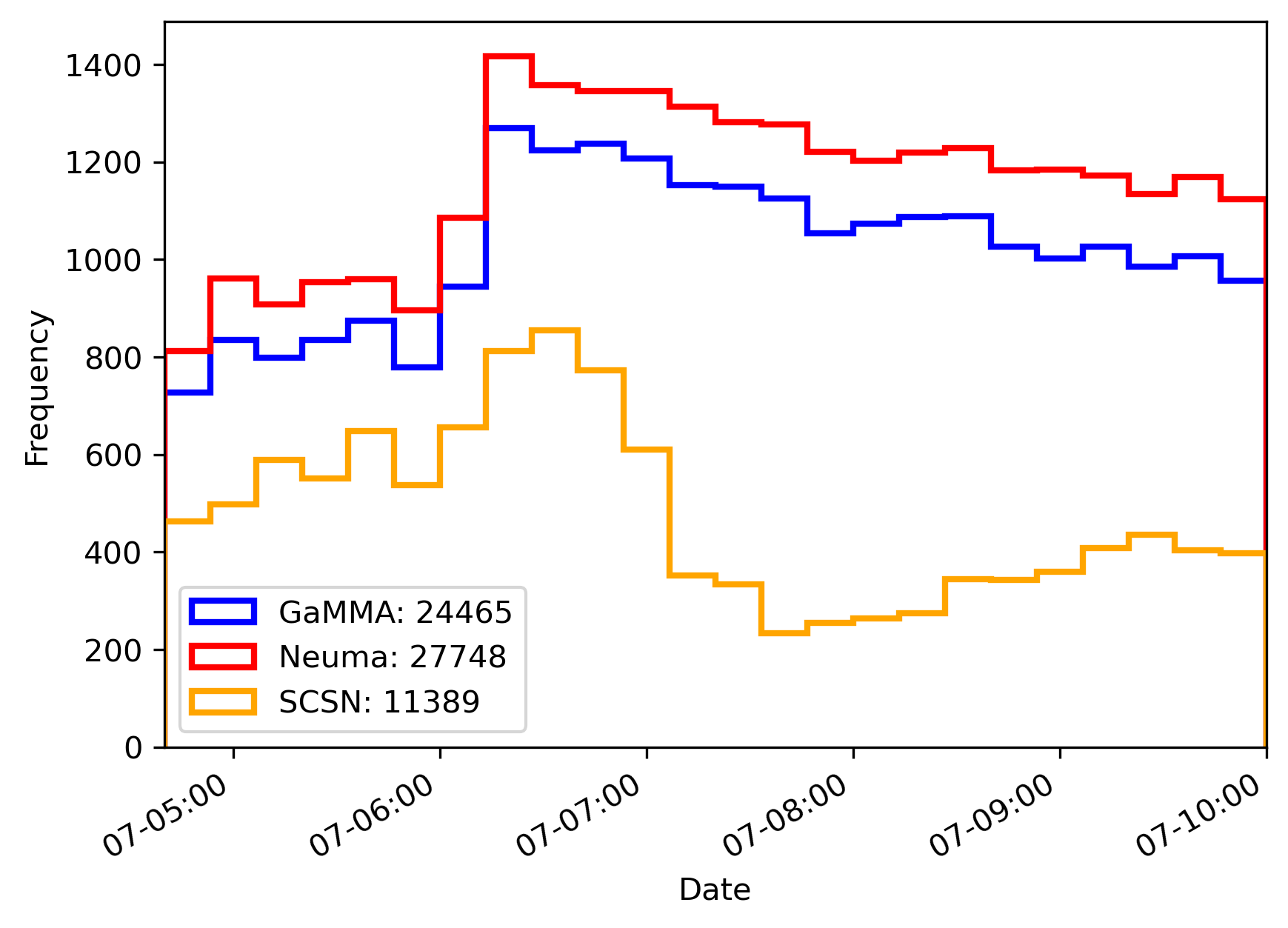}
    \label{fig:counttime}
    \end{subfigure}
        \caption{Summary of detection results over time. Left and right panels show magnitude and counts, respectively. Note the considerably larger drop in magnitude completeness after the M7.1 for the SCSN catalog than the Neuma and GaMMA catalogs.}
    \label{fig:dettime}
\end{figure}

\begin{figure}[htb]
    \centering
    \includegraphics[width=1.0\textwidth]{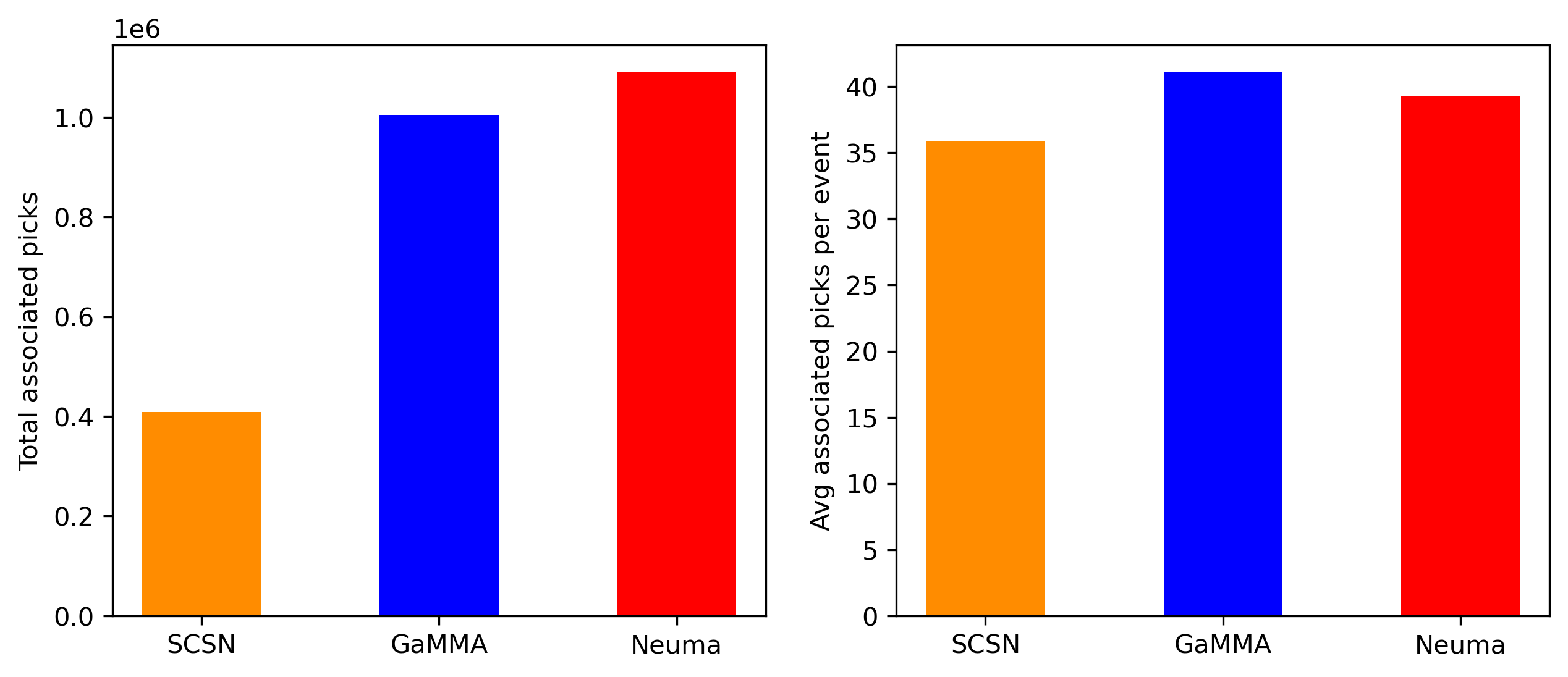}
    \caption{Ridgecrest dataset phase association results. }
    \label{fig:rcassoc}
\end{figure}

For this experiment, we set Neuma to output detections with at least 10 phase picks and set $b^T=0.35$~s and $b^A=1.0$. We use a 1D velocity model for southern California that is the same as the one employed by the SCSN for their real-time catalog \citep{hadley_seismic_1977}. The frequency-magnitude statistics for the three catalogs are shown in Figure \ref{fig:rcfreqmag}. Both GaMMA and Neuma detect more than twice as many earthquakes as the SCSN, which is almost a full magnitude unit decrease in the completeness magnitude due to the low b-value for this sequence. Neuma detects an additional 3283 events over GaMMA on this dataset. Additional gains of the algorithm are shown in Fig. \ref{fig:dettime}, where the results are parsed over time. Neuma detects more earthquakes than the other catalogs consistently over time. In comparison with the SCSN catalog, Neuma and GaMMA do not see such a large drop in the magnitude of completeness over the two days following the M 7.1 mainshock; these high-resolution catalogs have a much more gradual change in completeness over time.

We also examine the performance of Neuma at a phase association level. Figure \ref{fig:rcassoc} shows the phase association results for the three catalogs in two forms. Out of the 1,338,285 total picks made by PhaseNet, which represents an upper bound for the number of possible picks to be associated, Neuma associates 1,090,701 picks, while GaMMA only associates 1,005,348 -- a difference of about 85,000 picks. These numbers are much larger than the 409,115 associated by the SCSN. However, at an event level, Neuma associates almost 2 picks fewer per event than GaMMA. This decrease is expected because Neuma uses a much more appropriate model for the velocity and picking errors, which would result in GaMMA being more likely to include noise picks.

\begin{figure}[htb]
    \centering
    \includegraphics[width=1.0\textwidth]{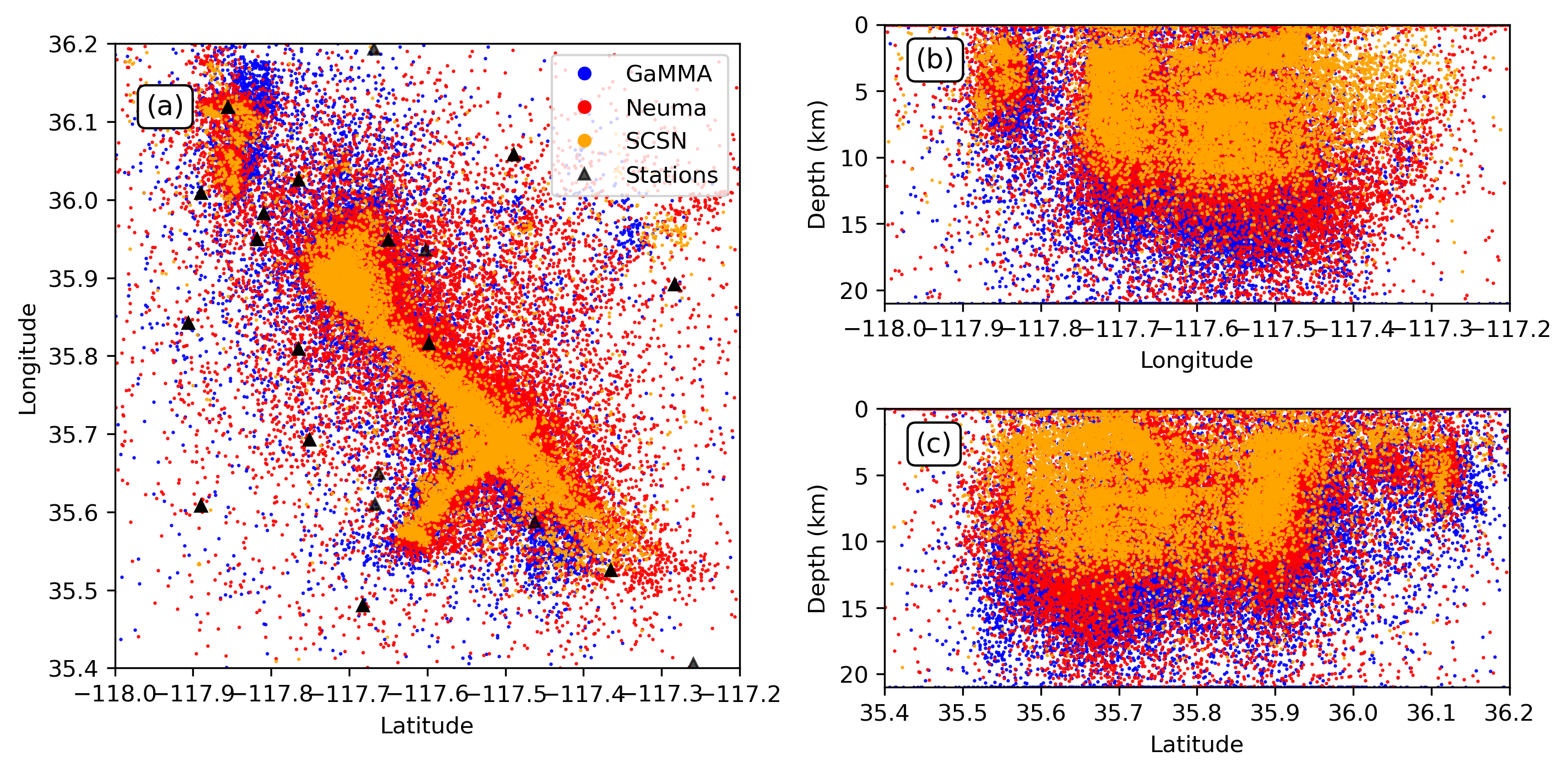}
    \caption{Comparison of hypocenters for the Ridgecrest dataset. }
    \label{fig:map}
\end{figure}

\begin{figure}[htb]
    \centering
    \includegraphics[width=0.65\textwidth]{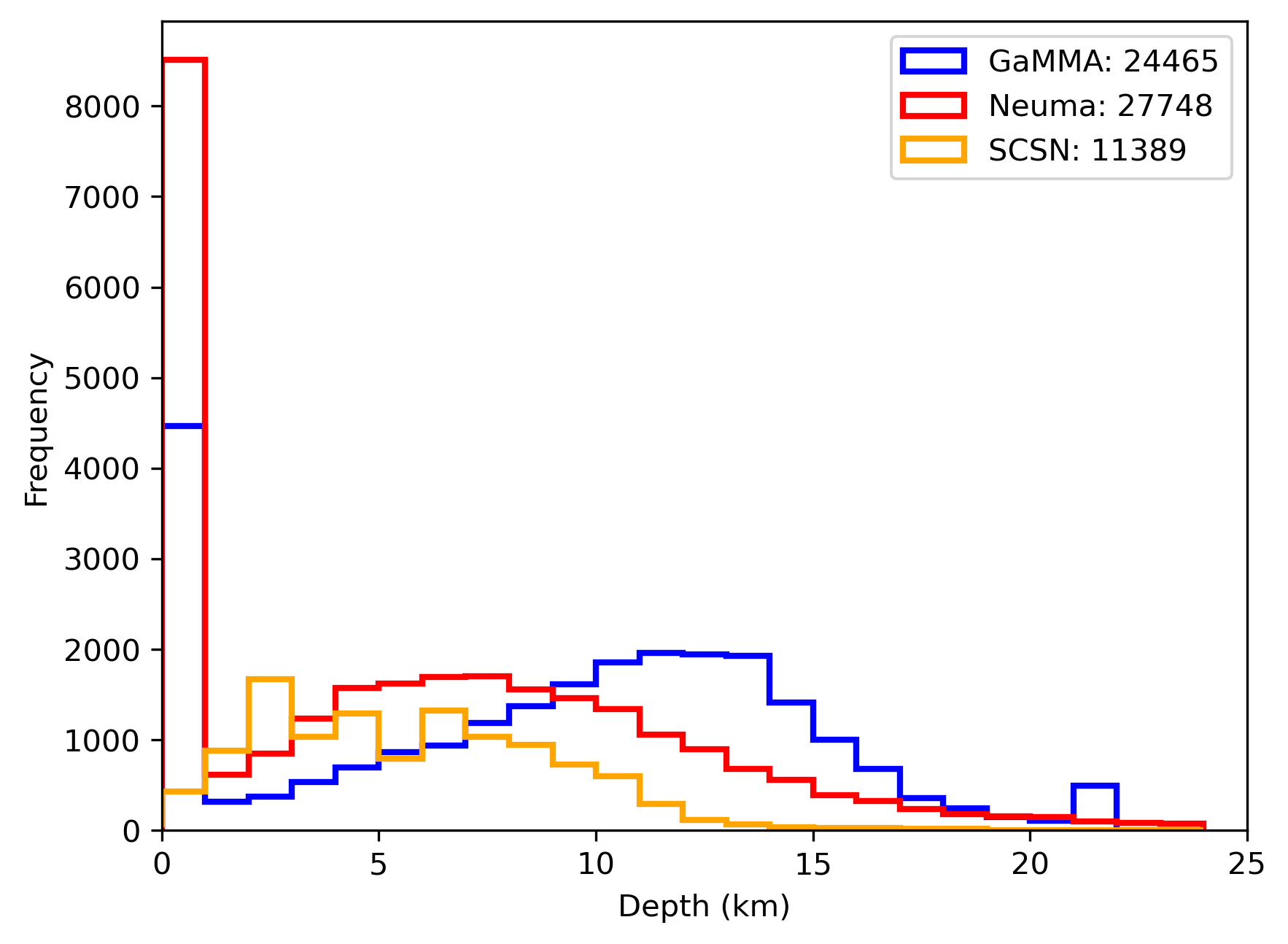}
    \caption{Depth distributions of earthquakes in each catalog.}
    \label{fig:depthdist}
\end{figure}

The more efficient association of Neuma results in locations that are substantially better quality. Figure \ref{fig:map} shows the hypocenters for the three catalogs. The SCSN results are the cleanest, which is expected because the picks have all been manually reviewed, and the detected events have the highest completeness magnitude, which results in a subset of events with the highest signal to noise ratio. Neuma locations in general closely track those of the SCSN, although with some minor differences expected to arise from differences in the velocity model used. GaMMA locations in contrast are significantly worse, with depths systematically more than 5 km deeper that reflects the large errors produced by the homogeneous velocity model and inclusion of more false positive associated picks.

\section{Conclusion}
We formulate a new mixture model-based phase association algorithm called Neuma, which incorporates a physics-informed deep neural network as the forward model to rapidly and reliable compute travel times in 3D media. Our formulation also explicitly includes a class for noise picks, which are widespread in automated picking datasets. Our formulation is easily implemented with the expectation-maximization algorithm by iterating between optimizing a set of hypocenters and magnitudes, and determining the optimal event to associate each pick to. We demonstrate significant improvements over several baseline methods on challenging synthetic and real datasets. The developed approach is expected to help streamline the automation of seismicity catalog construction.


\newpage

\bibliography{Arxiv}
\bibliographystyle{plainnat}

\end{document}